\documentclass[12pt]{article}
\usepackage{amssymb,amsmath,epsfig}
\allowdisplaybreaks

\begin{document}
\title{\bf Stability of the Closed Einstein Universe in Energy-Momentum Squared Gravity}
\author{M. Sharif \thanks {msharif.math@pu.edu.pk} and M. Zeeshan Gul
\thanks{mzeeshangul.math@gmail.com}\\
Department of Mathematics, University of the Punjab,\\
Quaid-e-Azam Campus, Lahore-54590, Pakistan.}
\date{}
\maketitle

\begin{abstract}
This paper analyzes the stability of the closed Einstein static
universe by using linear homogeneous perturbations in the framework
of energy-momentum squared gravity. This newly developed proposal
resolves the primordial singularity and yields feasible cosmological
results in the early times. For this purpose, we consider the closed
Friedmann-Robertson-Walker universe with isotropic matter
configuration and adopt small perturbations on the matter parameters
and scale factor. Further, we establish equations of motion for
static as well as perturbed systems and analyze the stable regions
for particular $f(\mathcal{R},\mathbf{T}^{2})$ models corresponding
to both conserved and non-conserved energy-momentum tensor. The
graphical interpretation demonstrates that stable regions of the
Einstein cosmos exist for all values of the model parameters and
equation of state variable. We conclude that stable regions in this
modified theory are large as compared to other modified theories of
gravity.
\end{abstract}

\section{Introduction}

The current cosmic acceleration has been the most stunning and
dazzling consequence for cosmologists over the past two decades.
Scientists claim that this acceleration is caused by an ambiguous
force dubbed as dark energy which has repulsive effects. This
mysterious force has motivated numerous researchers to reveal its
unknown attributes. In this framework, alternative gravitational
theories are regarded as the most important and optimistic proposals
to unveil the cryptic nature of dark components. These theories can
be developed by adding scalar invariants and their corresponding
functions in the geometric section of the Einstein-Hilbert action.
The simplest modification of general relativity (GR) is
$f(\mathcal{R})$ theory, where $\mathcal{R}$ is the Ricci scalar. A
comprehensive literature \cite{1} has been available to comprehend
the viable aspects of this theory.

The interactions among curvature and matter parts generalized
$f(\mathcal{R})$ theory. Such interactions determine the rotation
curves of galaxies as well as different cosmic stages. These
theories are non-conserved that provide the presence of an extra
force. These coupling proposals are very useful to comprehend the
cosmic acceleration and cryptic nature of dark components. Harko et
al. \cite{2} introduced such interactions in $f(\mathcal{R})$ theory
named $f(\mathcal{R},\mathcal{T})$ gravity, where $\mathcal{T}$ is
the trace of the energy-momentum tensor (EMT). The non-minimal
interaction between geometry and matter was developed in \cite{3}
named $f(\mathcal{R}, \mathcal{T},\mathcal{R}_{\alpha\beta}
\mathcal{T}^{\alpha\beta})$ theory. One of the modification gives
rise to $f(\mathbb{R}, \mathbb{T}^{\phi})$ theory, where $\phi$
defines the scalar field \cite{3a}.

The presence of singularities is considered a major problem in GR
because of its prediction at the high energy level, where GR is no
longer valid due to the expected quantum impacts. However, there is
no specific formalism for quantum gravity. In this regard, a new
covariant generalization of GR is developed by adding the analytic
function $\mathcal{T}^{\alpha\beta}\mathcal{T}_{\alpha\beta}$ in the
generic action known as energy-momentum squared gravity (EMSG). This
is also referred to as $f(\mathcal{R}, \mathbf{T}^{2})$ theory,
where $\mathbf{T}^{2}$ is denoted by
$\mathcal{T}^{\alpha\beta}\mathcal{T}_{\alpha\beta}$ \cite{4}. This
modification of GR is considered the most favorable and prosperous
technique which resolves the spacetime singularity in the
non-quantum description. Consequently, the corresponding field
equations are different from GR only in the presence of matter
sources. It contributes squared terms in the field equations that
are used to explore various fascinating cosmological consequences.

Board and Barrow \cite{5} studied the exact solutions of isotropic
spacetime and investigated their behavior with cosmic growth,
presence, and absence of singularities. Nari and Roshan \cite{7}
explored the feasible and stable compact stars in this context.
Bahamonde et al. \cite{8} analyzed the minimal and non-minimal EMSG
models and concluded that these models determine the cosmic
acceleration. Barbar et al \cite{9} investigated physically viable
conditions of the bouncing universe corresponding to a specific EMSG
model. Rudra and Pourhassan \cite{10} examined thermodynamic
characteristics of a black hole. Singh et al \cite{11} analyzed the
geometry of compact objects in the existence of quark matter.
Recently, we have discussed the Noether symmetry approach in this
framework and investigated the viable behavior of cosmological
solutions through different cosmological parameters. We have also
studied the dynamics and stability of dense objects \cite{12}. It is
found that modified EMSG terms boost the system's stability and
hence prevent the collapsing phenomenon. The above literature makes
it clear that this theory needs more focus and thus motivation to
explore this theory further is very much obvious.

It has been found that at early times, all matter was concentrated
to an infinitely compact point named primordial or big-bang
singularity. This singularity is considered a major problem in
cosmology. In this perspective, the emergent cosmic scenario is
identified as a useful technique that helps to resolve the
primordial singularity. According to this strategy, the cosmos
initiates from the static state then progresses to the expansion
process which causes the inflationary phase. In this context, the
early state of the cosmos is the Einstein universe (EU) rather than
primordial singularity. The presence, as well as stability of the EU
against all sorts of perturbations, justify the emergent cosmos.
Nevertheless, the concept of an emergent universe is not well
illustrated in GR due to the existence of unstable solutions.
Harrison \cite{13} examined the stable regions of the EU through
inhomogeneous perturbations in the presence of dust matter
configuration. It was examined that stable regions exist if
$5c^{2}_{s}>1$ is satisfied, where $c_{s}$ is the speed of sound
\cite{14}. Barrow and Yamamoto \cite{15} examined the stability of
EU corresponding to various matter configurations and obtained the
unstable regions against homogeneous perturbations.

Alternative gravitational theories have attained much attention to
find stable solutions for the EU. The stable solutions of the EU
employing homogeneous perturbations with specific $f(\mathcal{R})$
models have been analyzed in \cite{16}. In this framework, the
stable regions exist by homogenous perturbations whereas these
stable modes become unstable corresponding to inhomogeneous
perturbations. The stable modes of EU in $f(G)$ theory against
homogeneous perturbations has been examined in \cite{17}. The stable
regions of the EU by using scalar perturbations in modified
teleparallel theory have been discussed in \cite{18}. They found
that stable EU appears for both open and closed universe models. The
stable regimes in the modified Gauss-Bonnet theory against
homogenous and inhomogeneous perturbations have been studied in
\cite{19}. The stability of EU in scalar-field theories has been
investigated in \cite{20} and found the stable as well as unstable
regions corresponding to homogeneous and inhomogeneous
perturbations.

The analysis of stable EU has also been a source of fascinating
results in the context of curvature-matter interactions. The
stability of EU in $f(\mathcal{R},\mathcal{T})$ framework has been
explored in \cite{21} and found stable modes which were unstable in
$f(\mathcal{R})$ theory. The stable regions of EU corresponding to
both (\textit{homogeneous/inhomogeneous}) perturbations in
$f(G,\mathcal{T})$ theory has been analyzed in \cite{22}. The stable
modes of EU by using different perturbation techniques in
$f(\mathcal{R},\mathcal{T},\mathcal{R}_{\alpha\beta}\mathcal{T}^{\alpha\beta})$
background has been discussed in \cite{23}. They also studied the
stability criteria with inhomogeneous perturbations in
$f(\mathcal{R},\mathcal{T})$ gravity. The stable modes of EU with
anisotropic homogeneous perturbations in curvature-matter coupled
theory have been examined in \cite{24}.

In this paper, we use the homogeneous perturbations technique to
examine the stability of closed EU in EMSG. This analysis will help
to examine the impacts of curvature-matter interactions on the
stability of the closed EU. The paper is organized as follows.  We
derive the equations of motion in the background of the closed FRW
universe in section \textbf{2}. Section \textbf{3} investigates the
stable regions of the closed EU for conserved and non-conserved EMT.
A brief description and discussion of the outcomes are provided in
the last section.

\section{Einstein Static Universe}

This section formulates the equations of motion for closed FRW
spacetime in the context of $ f(\mathcal{R},\mathbf{T} ^{2})$
theory. The action for this gravity is \cite{4}
\begin{equation}\label{1}
\mathcal{S}=\frac{1}{2\kappa^2}\int f(\mathcal{R},\mathbf{T}
^{2})\sqrt{-g}d^{4}x+\int \mathcal{L}_{\textit{M}}\sqrt{-g}d^{4}x,
\end{equation}
where the coupling constant, determinant of the line element and
matter Lagrangian density is denoted by $\kappa^2$, $g$ and
$\mathcal{L}_{\textit{M}}$, respectively. This theory has maximum
energy density and correspondingly a minimum scale factor at the
early universe which indicates that there is a bounce at early
times.  Moreover, this theory possesses a true sequence of
cosmological eras. Although the cosmological constant does not play
a crucial role in the background of the standard cosmological model,
however, the cosmological constant supports resolving singularity
only after the matter-dominated era in EMSG. However, the profile of
density supports the inflationary cosmological models that
successfully provide convincing answers to major cosmological issues
like horizon problem, flatness problem, etc. It is worthwhile to
mention here that this theory overcomes the spacetime singularity
but does not change the cosmological evolution. It is assumed that
some useful results will be obtained to study the stability of the
closed EU due to the matter-dominated era. The following equations
of motion are obtained from the variation of action corresponding to
the metric tensor.
\begin{equation}\label{2}
\mathcal{R}_{\alpha\beta}f_{\mathcal{R}}+g_{\alpha \beta}\Box
f_{\mathcal{R}}-\nabla_{\alpha}\nabla_{\beta}f_{\mathcal{R}}
-\frac{1}{2}g_{\alpha\beta}f=\mathcal{T}_{\alpha\beta}
^{(\textit{M})} -\Theta_{\alpha\beta}f_{\mathbf{T}^{2}},
\end{equation}
where $\Box= \nabla_{\mu}\nabla^{\mu}$, $f\equiv
f(\mathcal{R},\mathbf{T}^{2})$, $f_{\mathbf{T}^{2}}= \frac{\partial
f} {\partial \mathbf{T}^{2}}$, $f_{\mathcal{R}}= \frac{\partial f}
{\partial \mathcal{R}}$,
\begin{eqnarray}\label{3}
\Theta_{\alpha\beta}=-2\mathcal{L}_{m}\left(\mathcal{T}
_{\alpha\beta}-\frac{1}{2}g_{\alpha\beta}\mathcal{T}
\right)-4\frac{\partial^{2}\mathcal{L}_{\textit{M}}}{\partial
g^{\alpha\beta}\partial g^{\mu\nu}}\mathcal{T}^{\mu\nu}
-\mathcal{T}\mathcal{T}_{\alpha\beta}+2\mathcal{T}
_{\alpha}^{\mu}\mathcal{T}_{\beta\mu},
\end{eqnarray}
This theory reduces to $f(\mathcal{R})$ gravity for $f(\mathcal{R},
\mathbf{T}^{2})= f(\mathcal{R})$ and to GR when $f(\mathcal{R},
\mathbf{T}^{2})= \mathcal{R}$. The stress-energy tensor demonstrates
the configuration of matter and energy in gravitational physics,
whereas every non-zero component provides dynamical parameters with
specific physical properties.

Here, we take into account isotropic matter configuration as
\begin{eqnarray}\label{4}
\mathcal{T}^{(\textit{M})}_{\alpha\beta}&=&\left(\mathrm{\rho}
+\mathrm{p}\right)\mathcal{U}_{\alpha}\mathcal{U}
_{\beta}-\mathrm{p} g_{\alpha\beta},
\end{eqnarray}
where four-velocity, matter pressure and density are defined by
$\mathcal{U}_{\alpha}$, $\mathrm{p}$ and $\mathrm{\rho}$,
respectively. Manipulating Eq.(\ref{3}), we obtain
\begin{equation}\nonumber
\Theta_{\alpha\beta}=-\left(3\mathrm{p}^{2}+\mathrm{\rho}^{2}
+4\mathrm{p}\mathrm{\rho}\right)\mathcal{U}_{\alpha}\mathcal{U}
_{\beta}.
\end{equation}
Rearranging Eq.(\ref{2}), we have
\begin{equation}\label{5}
\mathcal{G}_{\alpha\beta}=\frac{1}{f_{\mathcal{\mathcal{R}}}}
\left(\mathcal{T}_{\alpha\beta}^{(\textit{M})}
+\mathcal{T}_{\alpha\beta}^{(D)}\right),
\end{equation}
where $\mathcal{G}_{\alpha\beta}$ is the Einstein tensor and
$\mathcal{T}_{\alpha\beta}^{(D)}$ are the additional impacts of EMSG
that include the higher-order curvature terms because of the
modification in curvature part named dark source terms expressed as
\begin{eqnarray}\nonumber
\mathcal{T}_{\alpha\beta}^{(D)}=\frac{1}{2}
g_{\alpha\beta}\left(f-\mathcal{R}f_{\mathcal{R}}
\right)-\Theta_{\alpha \beta}f_{\mathbf{T}^{2}}
+\left(\nabla_{\alpha}\nabla_{\beta}-g_{
\alpha\beta}\Box\right)f_{\mathcal{R}}.
\end{eqnarray}
The $f(\mathcal{R},\mathbf{T}^{2})$ gravity provides non-conserved
EMT implying the presence of an extra force that acts as a
non-geodesic motion of particles given by
\begin{equation}\label{6}
\nabla^{\alpha}T^{(\textit{M})}_{\alpha\beta}
=-\frac{1}{2}\Big(f_{\mathbf{T}^{2}}
g_{\alpha\beta}\nabla^{\alpha}\mathbf{T}^{2}
-2\nabla^{\alpha}\left(f_{\mathbf{T}^{2}}
\Theta_{\alpha\beta}\right)\Big).
\end{equation}

In order to study the isotopic homogenous universe, we consider
closed FRW spacetime as \cite{16}
\begin{equation}\label{7}
ds^{2}= d\mathrm{t}^{2}-\frac{\mathrm{a}^{2}(\mathrm{t})}
{1-\mathrm{r}^{2}}d\mathrm{r}^{2}-\mathrm{a}^{2}(\mathrm{t})
\mathrm{r}^{2}d\theta^{2}-\mathrm{a}^{2}(\mathrm{t})
\mathrm{r}^{2}\sin^{2}{\theta}d\phi^{2},
\end{equation}
where $\mathrm{a}(\mathrm{t})$ is the scale parameter. The
corresponding field equations become
\begin{eqnarray}\nonumber
\frac{3}{\mathrm{a}^{2}}\left(1+\dot{\mathrm{a}}^{2}\right)&=&
\frac{1}{f_{\mathcal{R}}}\left\{\mathrm{\rho}+\frac{1}{2}
f\left(\mathcal{R},\mathbf{T}^{2}\right)+3\left(\frac{1
+\dot{\mathrm{a}}^{2}+\mathrm{a}\ddot{\mathrm{a}}}
{\mathrm{a}^{2}}\right)f_{\mathcal{R}}
\right.\\\label{8}&-&\left.\frac{3\dot{\mathrm{a}}}{\mathrm{a}}
\dot{f}_{\mathcal{R}}+\left(3\mathrm{p}^{2}+\mathrm{\rho}^{2}
+4\mathrm{p}\mathrm{\rho}\right)f_{\mathbf{T}^{2}}\right\},
\\\nonumber
-2\mathrm{a}\ddot{\mathrm{a}}-\left(1+\dot{\mathrm{a}}^{2}\right)&=&
\frac{1}{f_{\mathcal{R}}}\left\{\mathrm{a}^{2}\mathrm{p}
-\frac{\mathrm{a}^{2}}{2}f\left(\mathcal{R},\mathbf{T}^{2}\right)
-3\left(1+\dot{\mathrm{a}}^{2}+\mathrm{a}\ddot{\mathrm{a}}\right)
f_{\mathcal{R}} \right.\\\label{9}&+&\left.
2\mathrm{a}\dot{\mathrm{a}}\dot{f}_{\mathcal{R}}+\mathrm{a}^{2}
\ddot{f}_{\mathcal{R}}\right\}.
\end{eqnarray}
These equations describe how gravity and fluid parameters bend
spacetime. Here
\begin{eqnarray}\nonumber
\mathcal{R}=-6\left(\frac{\mathrm{a}\ddot{\mathrm{a}}
+\dot{\mathrm{a}}^{2}+1}{\mathrm{a}^{2}}\right),
\quad \mathbf{T}^{2}= 3\mathrm{p}^{2}+\mathrm{\rho}^{2},
\end{eqnarray}
where \textit{dot} demonstrates the derivative corresponding to
temporal coordinate. The non-conservation equation for perfect fluid
turns out to be
\begin{eqnarray}\nonumber
\dot{\mathrm{\rho}}+\frac{3\dot{\mathrm{a}}}{\mathrm{a}}
\left(\mathrm{\rho}+\mathrm{p}\right)&=&
\left(3\mathrm{p}^{2}+\mathrm{\rho}^{2}+4\mathrm{p}\mathrm
{\rho}\right)\dot{f}_{\mathbf{T}^{2}}-\frac{3\dot{\mathrm{a}}}
{\mathrm{a}}\left(3\mathrm{p}^{2}+\mathrm{\rho}^{2}+4\mathrm{p}
\mathrm{\rho}\right)f_{\mathbf{T}^{2}}
\\\label{10}
&-&f_{\mathbf{T}^{2}}\left\{\dot{\mathrm{\rho}}
\left(3\mathrm{\rho}+4\mathrm{p}\right)
+\dot{\mathrm{p}}\left(9\mathrm{p}+4\mathrm{\rho}\right)\right\}.
\end{eqnarray}

\section{Stability Analysis}

In this section, we examine the stability of closed EU by using
homogenous perturbations in EMSG. For this reason, the resulting
equations of motion minimize to
\begin{eqnarray}\label{11}
\frac{3}{\mathrm{a}_{0}^{2}}&=&\frac{1}{f_{\mathcal{R}}}
\left\{\mathrm{\rho}_{0}+\frac{1}{2}f\left(\mathcal{R}_{0},\mathbf{T}^{2}_{0}\right)
+\frac{3}{\mathrm{a}_{0}^{2}}f_{\mathcal{R}}+\left(3\mathrm{p}_{0}^{2}
+\mathrm{\rho}_{0}^{2}+4\mathrm{p}_{0}\rho_{0}\right)f_{\mathbf{T}^{2}}\right\},
\\\label{12}
-\frac{1}{\mathrm{a}_{0}^{2}}&=&\frac{1}{f_{\mathcal{R}}}\left\{\mathrm{p}_{0}
-\frac{1}{2}f\left(\mathcal{R}_{0},\mathbf{T}^{2}_{0}\right)
-\frac{3}{\mathrm{a}_{0}^{2}}f_{\mathcal{R}}+\ddot{f_{\mathcal{R}}}\right\},
\end{eqnarray}
where $\mathrm{a}_{0}$=constant,
$\mathcal{R}_{0}=\frac{6}{\mathrm{a}_{0}^{2}}=\mathcal{R}(\mathrm{a}_{0})$
and $\mathbf{T}_{0}^{2}=\mathrm{\rho}_{0}^{2}+3\mathrm{p}_{0}^{2}$.
Here $\mathrm{\rho}_{0}$ and $\mathrm{p}_{0}$ represent the
unperturbed matter variables. We take linear EoS
$(\mathrm{p}=\omega\mathrm{\rho})$ to examine the stable regions of
closed EU and determine the homogeneous perturbations in matter
density and scale parameter defined as
\begin{eqnarray}\label{13}
\mathrm{\rho}(\mathrm{t})=(\delta
\mathrm{\rho}(\mathrm{t})+1)\mathrm{\rho}_{0}, \quad
\mathrm{a}(\mathrm{t})=(\delta
\mathrm{a}(\mathrm{t})+1)\mathrm{a}_{0},
\end{eqnarray}
where perturbed matter density and scale parameter are represented
by $\delta \mathrm{\rho}(\mathrm{t})$ and $\delta
\mathrm{a}(\mathrm{t})$, respectively. By using Taylor series
expansion upto first order and consider
$f(\mathcal{R},\mathbf{T}^{2})$ as an analytic function, we have
\begin{equation}\label{14}
f(\mathcal{R},\mathbf{T}^{2})=
f(\mathcal{R}_{0},\mathbf{T}^{2}_{0})+f(\mathcal{R}_{0},\mathbf{T}^{2}_{0})\delta
\mathcal{R}+f(\mathcal{R}_{0},\mathbf{T}^{2}_{0})\delta
\mathbf{T}^{2},
\end{equation}
where $\delta \mathcal{R}$ and $\delta \mathbf{T}^{2}$ have the
following relations
\begin{eqnarray}\label{15}
\delta \mathcal{R}=-6\left(\delta \ddot{\mathrm{a}}-\frac{2\delta
\mathrm{a}}{\mathrm{a}^{2}_{0}}\right), \quad \delta \mathbf{T}^{2}=
\rho_{0}^{2}\left(3\omega^{2}+1\right)\delta \mathrm{\rho}.
\end{eqnarray}
Here $\delta
\ddot{\mathrm{a}}(\mathrm{t})=\frac{d^{2}}{d\mathrm{t}^{2}}(\delta
\mathrm{a}(\mathrm{t}))$. Using Eqs.(\ref{8})-(\ref{15}), the
linearized perturbed equations of motion turn out to be
\begin{eqnarray}\label{16}
&&6f_{\mathcal{R}}\delta
\mathrm{a}+\mathrm{a}_{0}^{2}\mathrm{\rho}_{0}\left\{1+\frac{\mathrm{\rho}_{0}}{2}
\left(15\omega^{2}+16\omega+5\right)f_{\mathbf{T}^{2}}\right\}\delta
\mathrm{\rho}=0,
\\\label{17}
&&2\left(\delta \ddot{\mathrm{a}}-\frac{1}{\mathrm{a}_{0}^{2}}\delta
\mathrm{a}\right)f_{\mathcal{R}}+\mathrm{\rho}_{0}\left\{\omega-\frac{1}{2}\mathrm{\rho}_{0}
\left(1+3\omega^{2}\right)f_{\mathbf{T}^{2}}\right\}\delta
\mathrm{\rho}=0.
\end{eqnarray}
These equations determine the direct relation between perturbed
matter density and scale factor.

The following sections address the stability of the closed EU for
vanishing and non-vanishing divergence of EMT.

\subsection{Stability for Conserved EMT}

It is well-known that curvature-matter coupled theories are
non-conserved. Here, we assume that the general conservation law
satisfies in EMSG and hence right side of Eq.(\ref{10}) must be zero
which yields
\begin{eqnarray}\nonumber
&&f_{\mathbf{T}^{2}}\left\{\dot{\mathrm{\rho}}\left(3\mathrm{\rho}
+4\mathrm{p}\right)+\dot{\mathrm{p}}\left(9\mathrm{p}+4\mathrm{\rho}
\right)\right\}+\frac{3\dot{\mathrm{a}}}{\mathrm{a}}
\left(3\mathrm{p}^{2}+\mathrm{\rho}^{2}+4\mathrm{p}\mathrm{\rho}
\right)f_{\mathbf{T}^{2}}
\\\label{18}
&&-\left(3\mathrm{p}^{2}+\mathrm{\rho}^{2}+4\mathrm{p}
\mathrm{\rho}\right)\dot{f}_{\mathbf{T}^{2}}=0.
\end{eqnarray}
We use Eqs.(\ref{16}) and (\ref{17}) to obtain the perturbed
equation of motion in terms of perturbed scale factor as follows
\begin{eqnarray}\nonumber
&&\delta \mathrm{a}\left\{3\omega+1+\frac{\mathrm{\rho}_{0}}{2}
\left(6\omega^{2}+16\omega+2\right)f_{\mathbf{T}^{2}}\right\}-\delta
\ddot{\mathrm{a}}
\\\label{19}
&&\times\mathrm{a}_{0}^{2}\left\{1+\frac{\mathrm{\rho}_{0}}{2}
\left(15\omega^{2}+16\omega+5\right)f_{\mathbf{T}^{2}}\right\}=0.
\end{eqnarray}
Manipulating Eqs.(\ref{11}) and (\ref{12}), we have
\begin{equation}\label{20}
\mathrm{a}_{0}^{2}=\frac{2f_{\mathcal{R}}}{\mathrm{\rho}_{0}\left(1+\omega\right)
+\mathrm{\rho}^{2}_{0}\left(3\omega^{2}+4\omega+1\right)f_{\mathbf{T}^{2}}}.
\end{equation}
Using this value in Eqs.(\ref{19}), the perturbed equation of motion
becomes
\begin{eqnarray}\nonumber
&&\delta \mathrm{a}\left\{3\omega+1+\frac{\mathrm{\rho}_{0}}{2}
\left(6\omega^{2}+16\omega+2\right)f_{\mathbf{T}^{2}}\right\}
\left\{\mathrm{\rho}^{2}_{0}\left(3\omega^{2}+4\omega+1\right)f_{\mathbf{T}^{2}}
\right.\\\label{21}
&&+\left.\mathrm{\rho}_{0}\left(1+\omega\right)\right\}-f_{\mathcal{R}}
\left\{2+\mathrm{\rho}_{0}\left(15\omega^{2}+16\omega+5\right)f_{\mathbf{T}^{2}}
\right\}\delta\ddot{\mathrm{a}}=0.
\end{eqnarray}
We also have fourth-order perturbed equations of motion as appearing
in other modified theories but only first-order linear terms appear
due to the existence of $\mathbf{T}^{2}$ term. Therefore, the
second-order perturbed field equation is obtained about the scale
factor in this gravity. Equation (\ref{21}) minimizes to the
following desired form in the GR limits ($f_{\mathcal{R}}=1$ and
$f_{\mathbf{T}^{2}}=0$) as
\begin{equation}\nonumber
2\delta \ddot{\mathrm{a}}-\mathrm{\rho}(3\omega^{2}+4\omega+1)\delta
\mathrm{a}=0.
\end{equation}
The solution of the perturbed field equation (21) helps to
investigate the stable regions of the closed Einstein universe but
this equation is quite complicated due to the presence of
multivariate functions and their derivatives.

To explore the viability of this result in EMSG, we consider a
specific type of generic function which gives minimal and
non-minimal couplings between geometric and matter parts. The
non-minimal model, $f(\mathcal{R},
\mathbf{T}^{2})$=$f_{0}\mathcal{R}\mathbf{T}^{2}$, yields the
perturbed field equation in a complex form and we are unable to
conclude any outcome from this model. So, for the sake of
convenience, we take the minimal coupling model as \cite{4}
\begin{equation}\label{22}
f(\mathcal{R},\mathbf{T}^{2})=f_{1}(\mathcal{R})+f_{2}(\mathbf{T}^{2}).
\end{equation}
This model represents an interesting cosmological model which can
explain the current evolution of the universe and the emergence of
the accelerated expansion as a geometrical consequence. We have
considered conserved EMT, hence the conservation law satisfies this
model.

Manipulating Eq.(\ref{18}), we obtain
\begin{eqnarray}\nonumber
f_{2}'(\mathbf{T}^{2})\left(9\omega^{2}+8\omega+3\right)
+f_{2}''(\mathbf{T}^{2})\left(3\omega^{2}+4\omega+1\right)
\left(6\omega^{2}+2\right)\mathrm{\rho}_{0}^{2}=0,
\end{eqnarray}
where prime determines the derivative corresponding to $x$, i.e.,
$x=\mathcal{R}$ or $\mathbf{T}^{2}$. The solution of the above
equation is given by
\begin{eqnarray}\label{23}
f_{2}(\mathbf{T}^{2})=b_{1}e^{\frac{-\left(9\omega^{2}+8\omega+3\right)
\mathbf{T}^{2}}{2\mathrm{\rho}_{0}^{2}\left(\omega+1\right)
\left(3\omega+1\right)\left(3\omega^{2}+1\right)}}+b_{2},
\end{eqnarray}
where $b_{1}$ and $b_{2}$ are constants of integration. It should be
noted that only this expression of $f(\mathbf{T}^{2})$ satisfies the
conservation law in the model (\ref{22}). Now using the preceding
equations, Eq.(\ref{21}) turns out to be
\begin{eqnarray}\label{24}
\left(\Delta_{1}\Delta_{3}-\Delta_{1}\Delta_{4}\Delta_{7}-\Delta_{2}\Delta_{3}\Delta_{7}
+\Delta_{2}\Delta_{4}\Delta_{7}^{2}\right)\delta \mathrm{a}
-f'_{1}(\mathcal{R})\left(\Delta_{5}-\Delta_{6}\Delta_{7}\right)\delta
\ddot{\mathrm{a}}=0,
\end{eqnarray}
where $\Delta_{i}'s (i=1,2,3,4,5,6,7)$ are
\begin{eqnarray}\nonumber
\Delta_{1}&=&
4\mathrm{\rho}_{0}^{2}\left(9\omega^{2}+6\omega
+1\right)\left(\omega+1\right)\left(3\omega^{2}+1\right),
\\\nonumber
\Delta_{2}&=&\mathrm{\rho}_{0}b_{1}\left(6\omega^{2}
+16\omega+2\right)\left(9\omega^{2}+8\omega+3\right),
\\\nonumber
\Delta_{3}&=&
2\mathrm{\rho}_{0}^{3}\left(\omega^{2}+2\omega+1\right)
\left(3\omega+1\right)\left(3\omega^{2}+1\right),
\\\nonumber
\Delta_{4}&=&\mathrm{\rho}_{0}^{2}b_{1}\left(3\omega^{2}
+4\omega+1\right)\left(9\omega^{2}+8\omega+3\right),
\\\nonumber
\Delta_{5}&=&
8\mathrm{\rho}_{0}^{2}\left(\omega+1\right)
\left(3\omega+1\right)\left(3\omega^{2}+1\right),
\\\nonumber
\Delta_{6}&=&2\mathrm{\rho}_{0}b_{1}\left(15\omega^{2}
+16\omega+5\right)\left(9\omega^{2}+8\omega+3\right),
\\\nonumber
\Delta_{7}&=& e^{\frac{-\left(9\omega^{2}+8\omega+3\right)
}{2\mathrm{\rho}_{0}^{2}\left(\omega+1\right)\left(3\omega
+1\right)\left(3\omega^{2}+1\right)}}.
\end{eqnarray}
Equation.(\ref{24}) has the following solution
\begin{equation}\nonumber
\delta a(t)=c_{1}e^{\Omega t}+c_{2}e^{-\Omega t},
\end{equation}
where integration constants are represented by $c_{1}$ and $c_{2}$
and the factor $\Omega$ defines the frequency of small perturbation
given by
\begin{equation}\label{25}
\Omega^{2}=\frac{\left(\Delta_{3}-\Delta_{4}\Delta_{7}\right)
\left(\Delta_{1}-\Delta_{2}\Delta_{7}\right)}
{f'_{1}(\mathcal{R})\left(\Delta_{5}-\Delta_{6}\Delta_{7}\right)}.
\end{equation}
This parameter must be less than zero to prevent the exponential
growth or collapse which yields stability of the closed EU. This
frequency in GR limit is defined as
\begin{equation}\nonumber
\Omega^{2}=\frac{\rho_{0}}{2}(3\omega^{2}+4\omega+1),
\end{equation}
which yields stable modes in the interval
$\omega\in(-1,-\frac{1}{3})$ \cite{17}.

We consider $\rho_{0}=0.3$ \cite{25} and
$f'_{1}(\mathcal{R})=\xi_{1}$ as a new parameter to examine the
stable modes of closed EU graphically. Here, the red regions
represent the stable modes of the closed EU. The graphical
interpretation in Figure \textbf{1} determines the stable regions of
closed EU corresponding to $\Omega^{2}$ against homogeneous
perturbations for $b_{1}>0$. It is analyzed that stable regions are
obtained for all values of $\omega$ and these stable modes become
more smooth as the integration constant decreases. Figure \textbf{2}
shows the stability of closed EU for negative values of $b_{1}$. It
is examined that stable regions become more smooth corresponding to
$\omega$ as the value of $b_{1}$ increases. It is clear from Figures
\textbf{1} and \textbf{2} that we obtain the stable regions of
closed EU which were not stable in other modified theories of
gravity \cite{22}-\cite{24}.
\begin{figure}
\epsfig{file=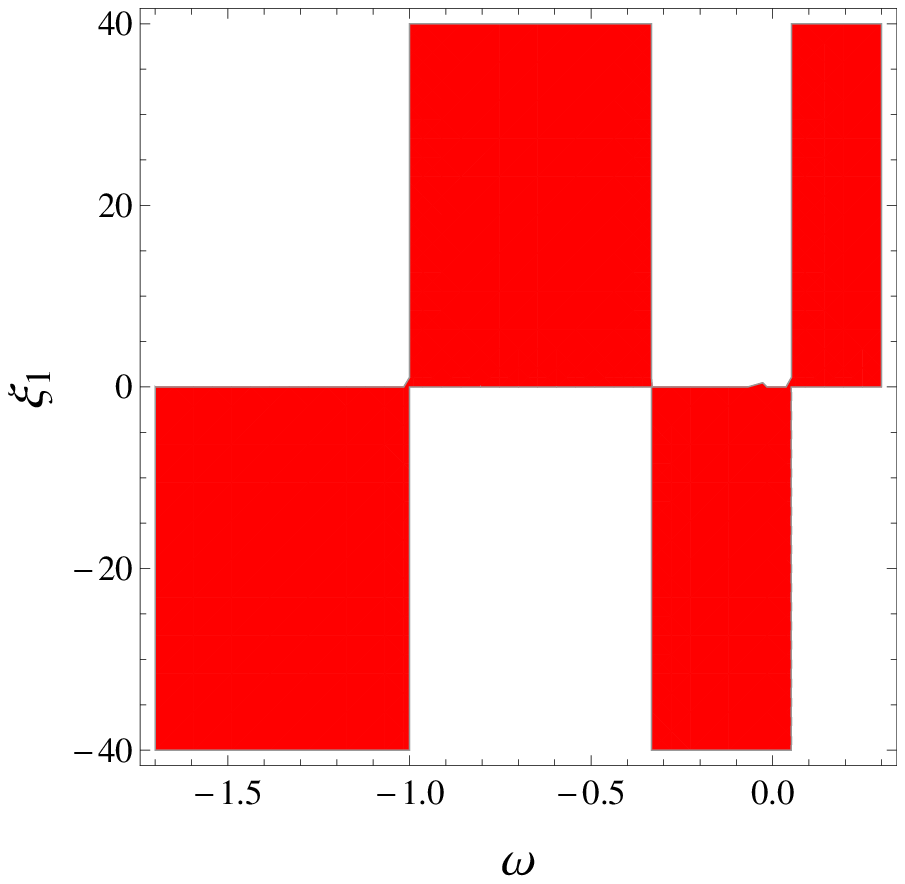,width=.5\linewidth}
\epsfig{file=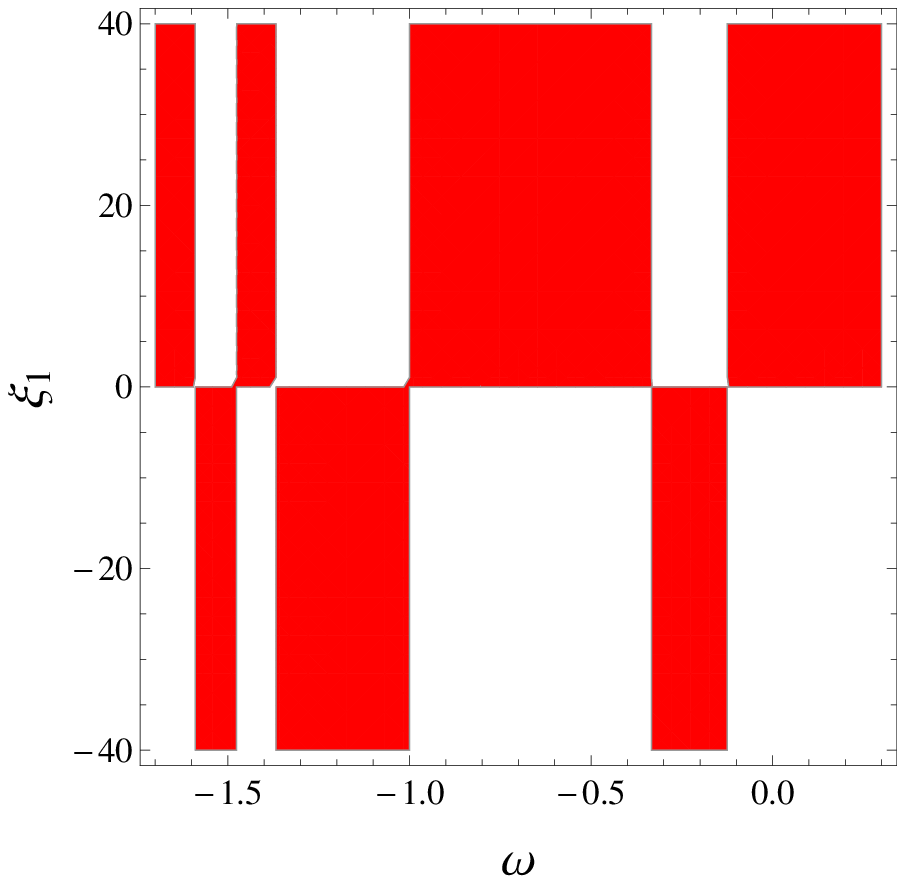,width=.5\linewidth}
\epsfig{file=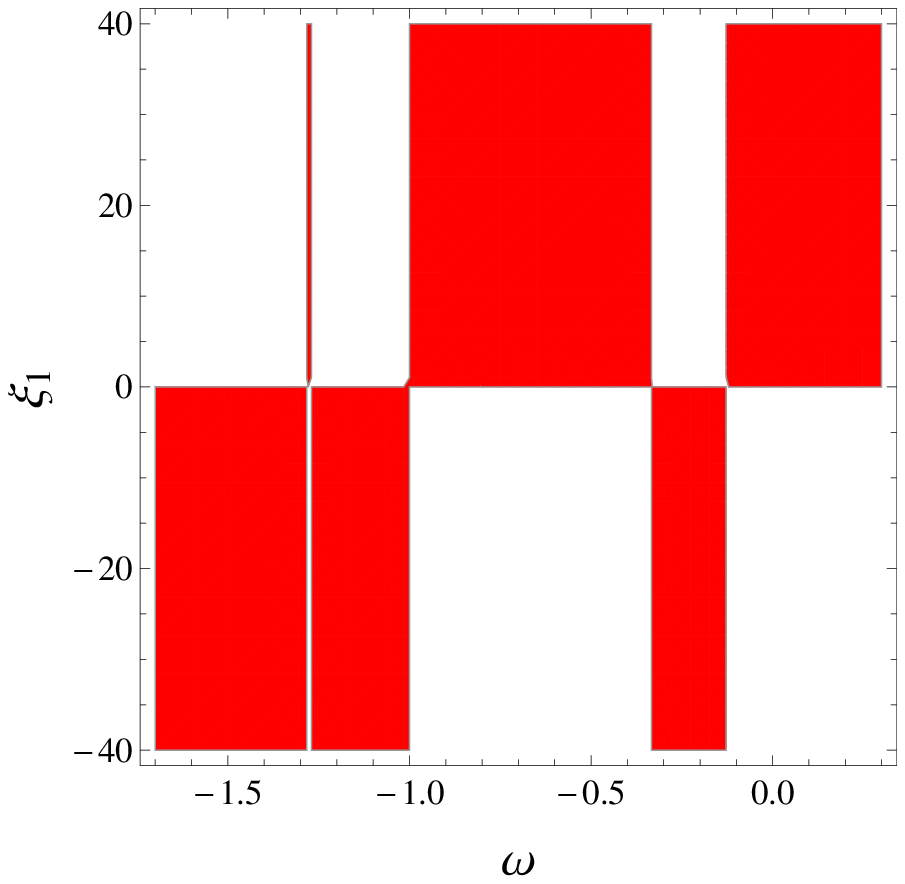,width=.5\linewidth}
\epsfig{file=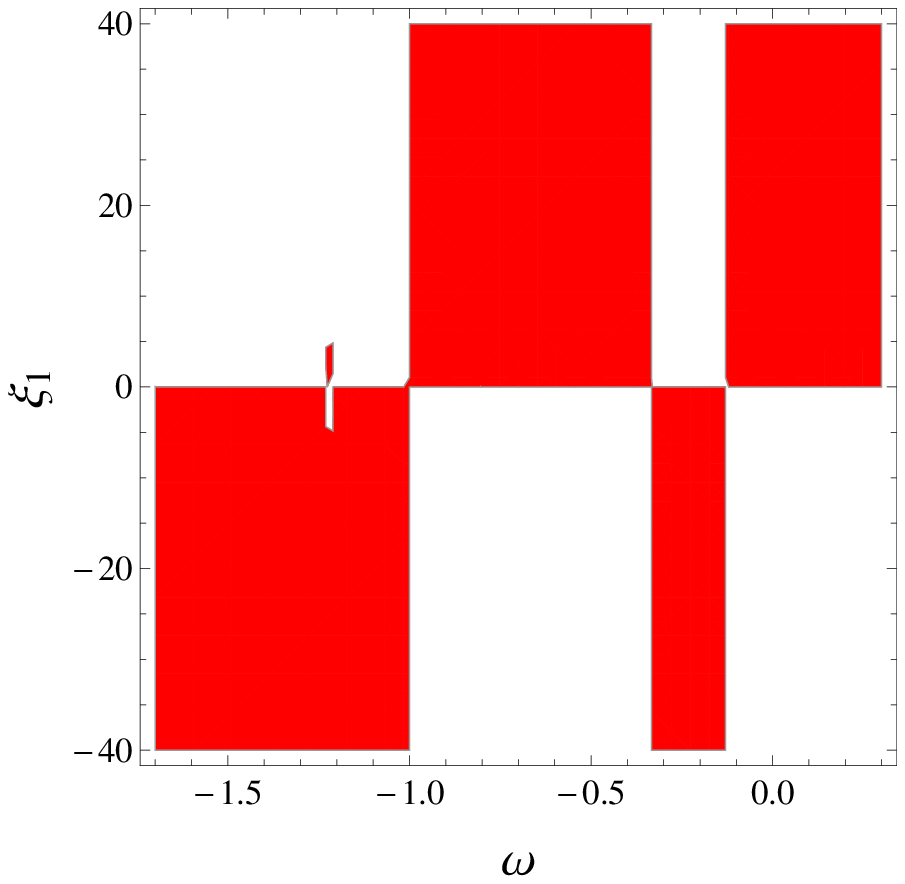,width=.5\linewidth} \caption{Stable modes in
$(\omega, \xi_{1})$ space corresponding to $b_{1}$=1,15,30 and 60
from left to right.}
\end{figure}
\begin{figure}
\epsfig{file=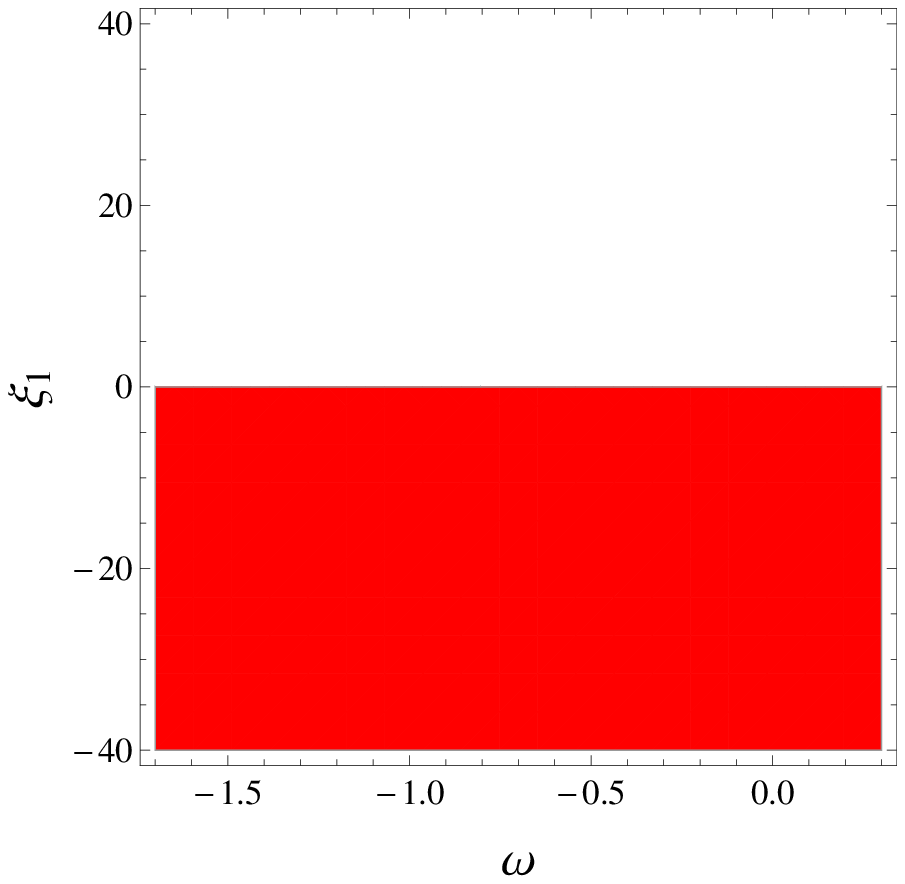,width=.5\linewidth}
\epsfig{file=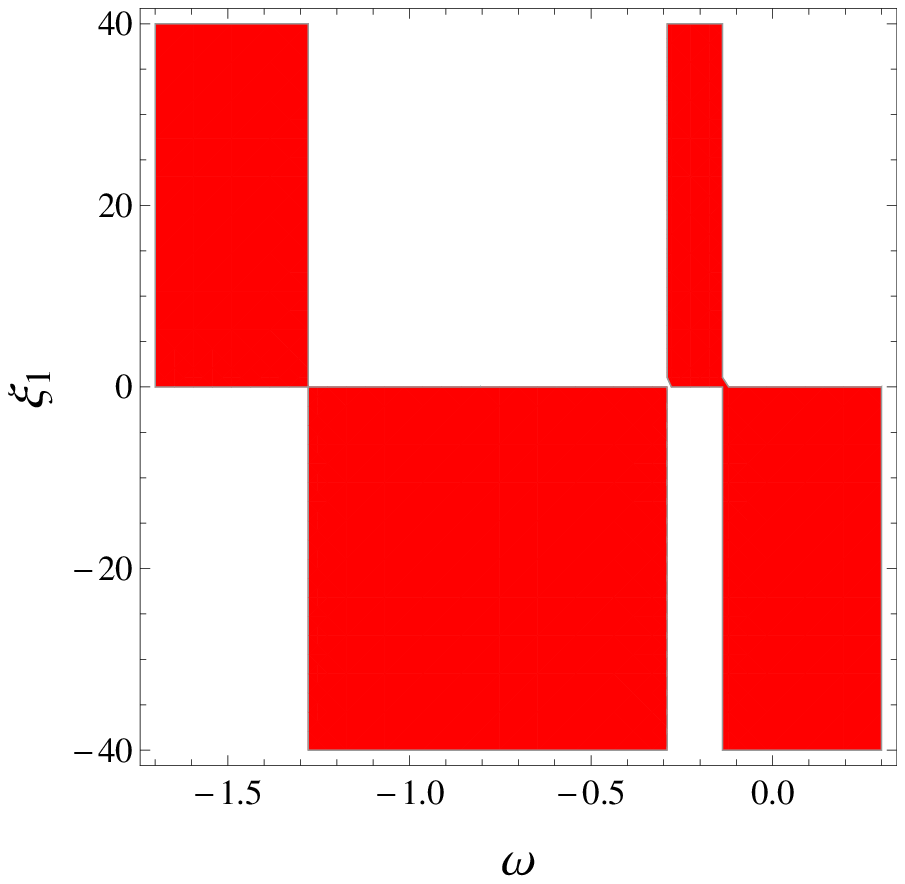,width=.5\linewidth}
\epsfig{file=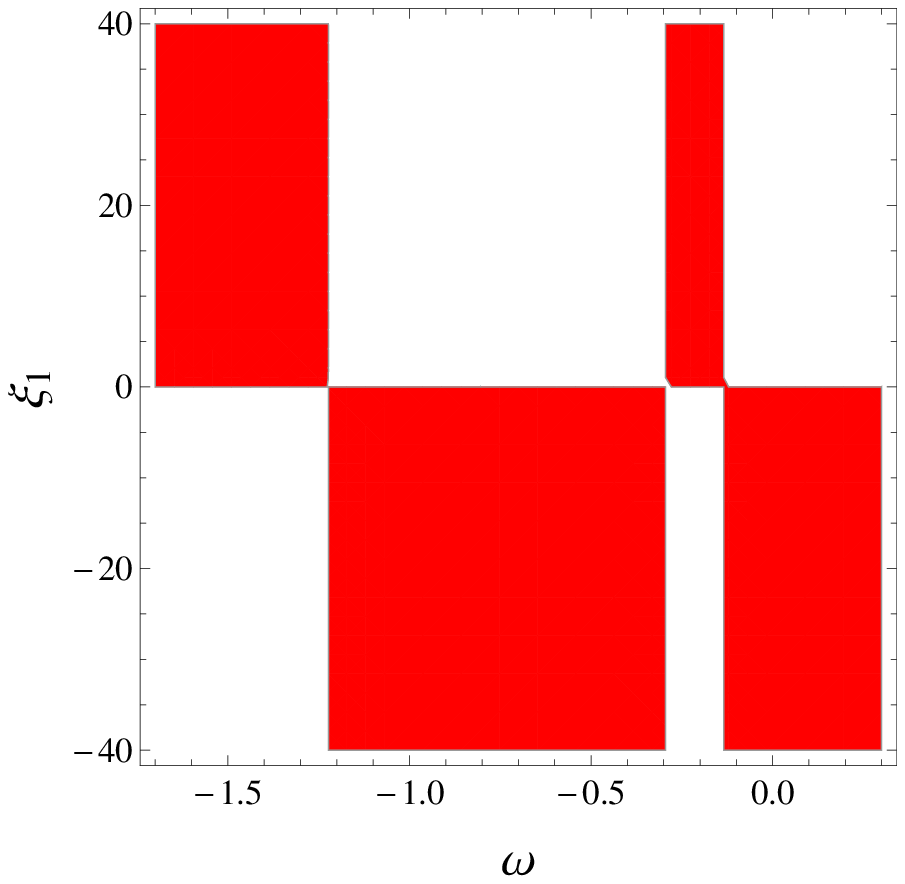,width=.5\linewidth}
\epsfig{file=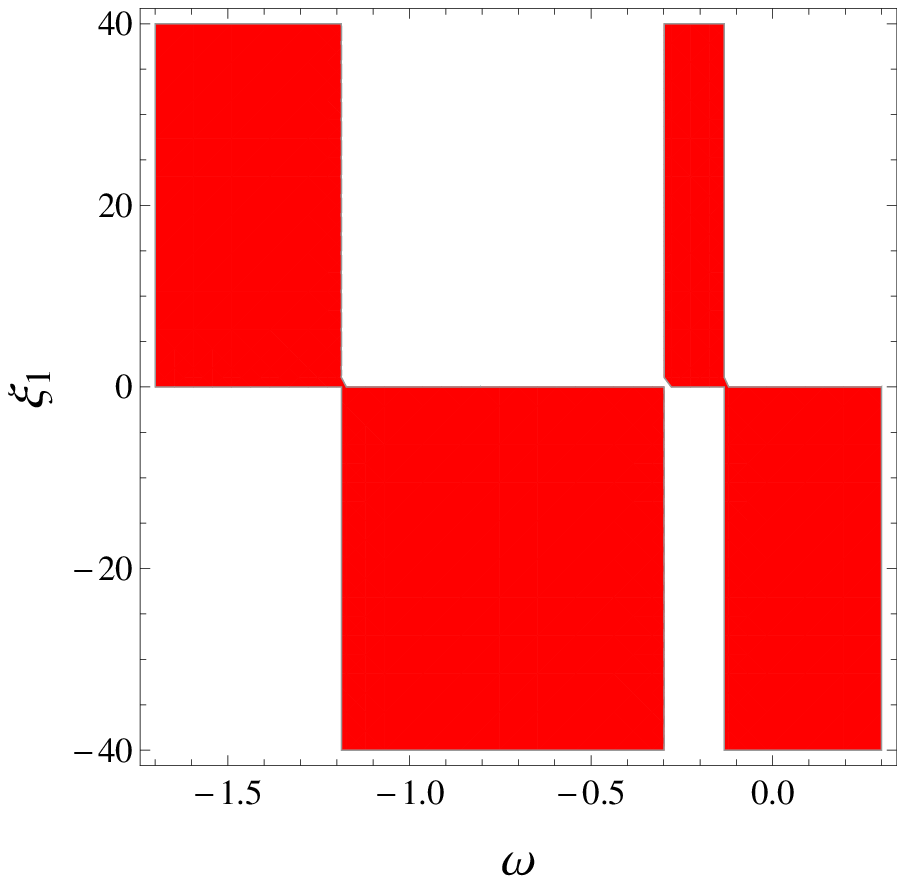,width=.5\linewidth}\caption{Stable modes in
$(\omega, \xi_{1})$ space corresponding to $b_{1}$=-1,-15,-30 and
-60 from left to right.}
\end{figure}

\subsection{Stability for Non-Conserved EMT}

This section analyzes stable regions of the closed EU for non-zero
divergence of EMT. Equation (\ref{6}) determines the general
expression of non-conserved EMT, which comes because of the
higher-order derivatives of EMT that naturally exist in the
equations of motion. This appears to be a problem for all
alternative theories of gravity that include the interaction between
curvature and matter parts. We consider a specific EMSG model as
\cite{5}
\begin{equation}\label{26}
f(\mathcal{R},\mathbf{T}^{2})=f_{1}(\mathcal{R})+\chi(\mathbf{T}^{2})^{n},
\end{equation}
to explore the impacts of non-zero divergence of EMT on the
stability of closed EU. Here, $\chi$ is the arbitrary constant. The
perturbed equations of motion for the considered model turn out to
be
\begin{eqnarray}\nonumber
&&\mathrm{a}_{0}^{2}\left\{\mathrm{\rho}_{0}+\frac{\chi
n}{2}\left(\mathrm{\rho}_{0}^{2}\left(3\omega^{2}+1\right)
\right)^{n-1}\left(15\omega^{2}+16\omega+5\right)
\mathrm{\rho}_{0}^{2}\right\}\delta\mathrm{\rho}
\\\label{27}
&&+6f_{1}'(\mathcal{R})\delta \mathrm{a}=0,
\\\nonumber
&&\left\{\rho_{0}\omega-\frac{\chi n}{2}
\left(\mathrm{\rho}_{0}^{2}\left(3\omega^{2}+1\right)
\right)^{n-1}\left(1+3\omega^{2}\right)\mathrm{\rho}_{0}^{2}
\right\}\delta\mathrm{\rho}
\\\label{28}
&&+ 2\left(\delta
\ddot{\mathrm{a}}-\frac{1}{\mathrm{a}_{0}^{2}}\delta
\mathrm{a}\right)f_{1}'(\mathcal{R})=0.
\end{eqnarray}
These equations determine the correlation between perturbed matter
density as well as the scale factor. The perturbed equation of
motion in terms of perturbed scale parameter becomes
\begin{eqnarray}\nonumber
&&\delta \mathrm{a}\left\{3\omega+1+\frac{\chi n}{2}
\left(\mathrm{\rho}_{0}^{2}\left(3\omega^{2}+1\right)\right)^{n-1}
\left(6\omega^{2}+16\omega+2\right)\mathrm{\rho}_{0}\right\}-\delta
\ddot{\mathrm{a}}
\\\label{29}
&&\times \mathrm{a}_{0}^{2}\left\{1+\frac{\chi
n}{2}\left(\mathrm{\rho}_{0}^{2}\left(3\omega^{2}+1\right)
\right)^{n-1}\left(15\omega^{2}+16\omega+5\right)\rho_{0}\right\}=0.
\end{eqnarray}
The addition of Eqs.(\ref{11}) and (\ref{12}) with respect to the
model (\ref{26}) gives
\begin{equation}\nonumber
a_{0}^{2}=2f'_{1}(\mathcal{R})\left\{\mathrm{\rho}_{0}
\left(\left(1+\omega\right)+n\chi\left(\mathrm{\rho}_{0}^{2}
\left(3\omega^{2}+1\right)\right)^{n-1}\left(3\omega^{2}
+4\omega+1\right)\mathrm{\rho}_{0}^{2}\right)\right\}^{-1}.
\end{equation}
Substituting this value in Eq.(\ref{29}), the corresponding
differential equation becomes
\begin{eqnarray}\nonumber
&&\delta \mathrm{a}\left(\Delta_{1}+\rho_{0}^{2n-2}\chi
n\Delta_{2}\right)\left(\Delta_{3}+\rho_{0}^{2n-2}\chi
n\Delta_{4}\right)-f'_{1}(\mathcal{R})\rho_{0}\delta
\ddot{\mathrm{a}}
\\\label{30}
&&\left(\frac{2}{\rho_{0}}+\rho_{0}^{2n-2}\chi
n\Delta_{5}\right)=0,
\end{eqnarray}
where $\Delta_{i}'s (i=1,2,3,4)$ are
\begin{eqnarray}\nonumber
\Delta_{1}&=& 3\omega+1,
\\\nonumber
\Delta_{2}&=&\frac{\mathrm{\rho}_{0}}{2}
\left(6\omega^{2}+16\omega+2\right)\left(\left
(3\omega^{2}+1\right)\right)^{n-1},
\\\nonumber
\Delta_{3}&=& \rho_{0}\left(1+\omega\right),
\\\nonumber
\Delta_{4}&=&\mathrm{\rho}^{2}_{0}\left(3\omega^{2}
+4\omega+1\right)\left(\left(3\omega^{2}+1\right)\right)^{n-1},
\\\nonumber
\Delta_{5}&=& \left(15\omega^{2}+16\omega+5\right)
\left(\left(3\omega^{2}+1\right)\right)^{n-1}.
\end{eqnarray}
Equation (\ref{30}) yields the following solution
\begin{equation}\nonumber
\delta \mathrm{a}(\mathrm{t})=d_{1}e^{\Omega t}+d_{2}e^{-\Omega
\mathrm{t}},
\end{equation}
where
\begin{equation}\nonumber
\Omega^{2}=\frac{\Delta_{4}\mathrm{\rho}_{0}n\chi
\left(\Delta_{1}+\Delta_{2}\chi
n\mathrm{\rho}_{0}^{n}\right)\Delta_{3}\mathrm
{\rho}_{0}\left(\Delta_{2}\chi
n+\Delta_{1}\mathrm{\rho}_{0}^{-2(n-1)}\right)}
{\xi_{2}\Delta_{5}\chi n}.
\end{equation}
Here $\xi_{2}=\rho_{0}f'_{1}(\mathcal{R})$.

The graphical behavior of closed EU stability for $n=1$ and
different values of the model parameters is given in Figures
\textbf{3} and \textbf{4}. It is examined that stable modes appear
for all values of the model parameters. Figures \textbf{5} and
\textbf{6} show the stability of closed EU for $\chi=1$ and distinct
values of $n$. We find that stable regions appear for all values of
the EoS variable and these stable modes become more smooth as $n$
increases. It is observed that the obtained stable regions are
unstable in other alternative gravitational theories.
\begin{figure}
\epsfig{file=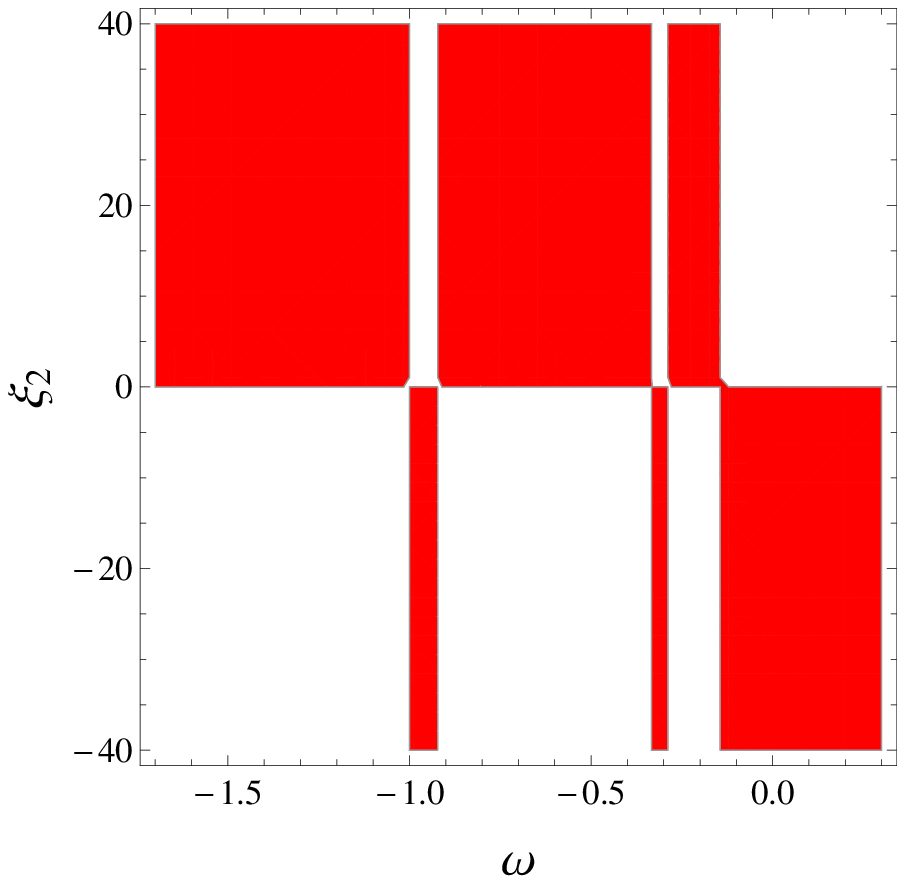,width=.5\linewidth}
\epsfig{file=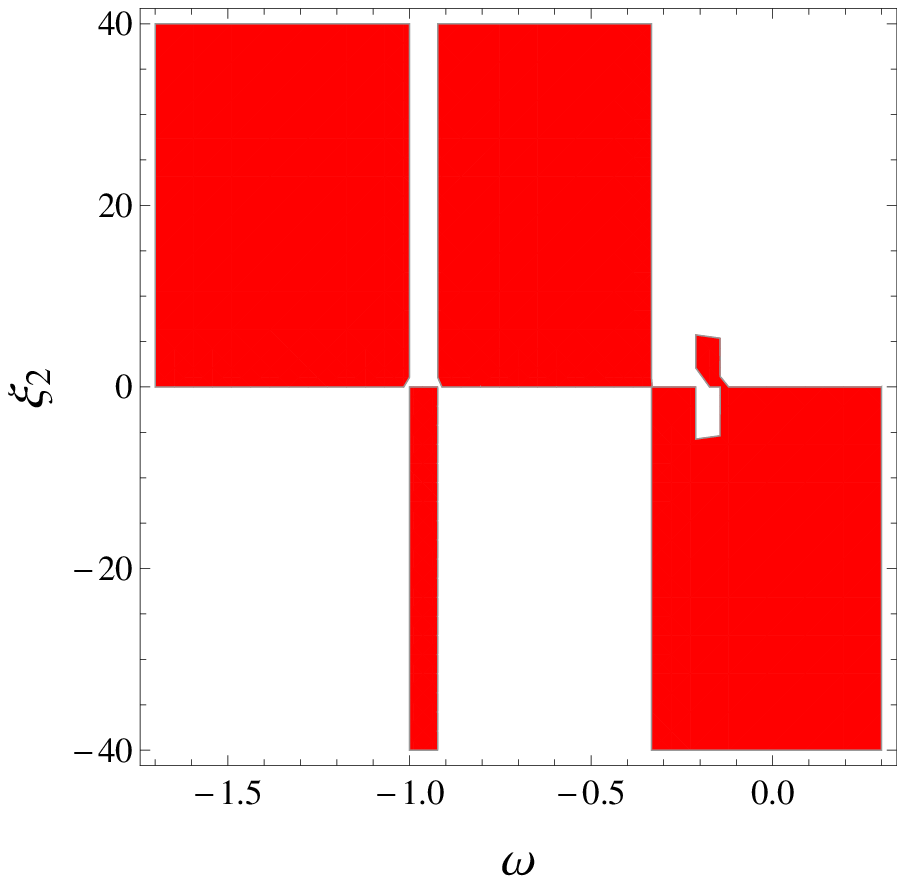,width=.5\linewidth}\caption{Stable modes in
$(\omega, \xi_{2})$ space corresponding to $\chi$=1,5 from left to
right.}
\end{figure}
\begin{figure}
\epsfig{file=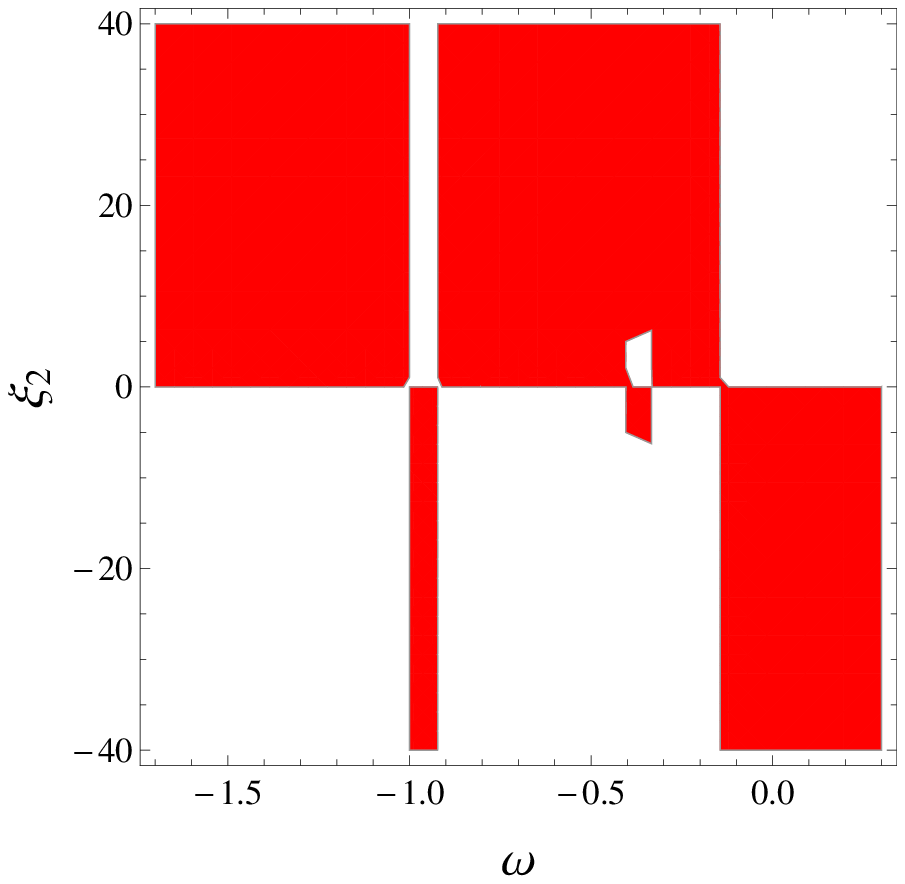,width=.5\linewidth}
\epsfig{file=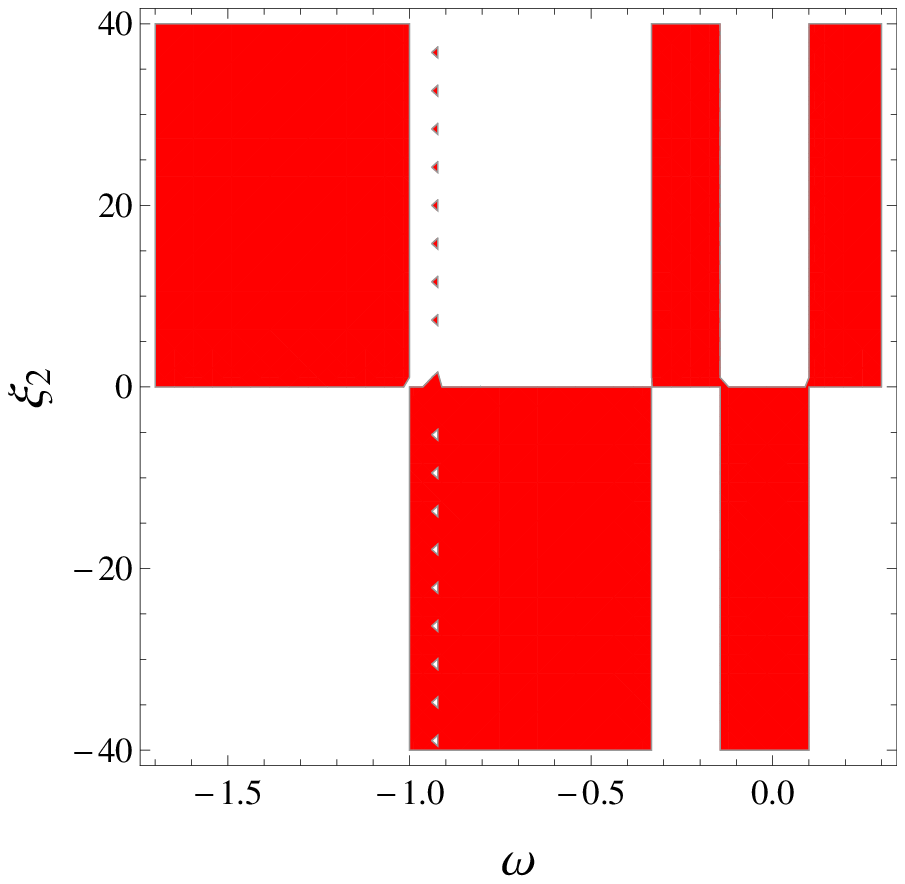,width=.5\linewidth} \caption{Stable regions in
$(\omega, \xi_{2})$ space corresponding to $\chi$=-1,-5 from left to
right.}
\end{figure}
\begin{figure}
\epsfig{file=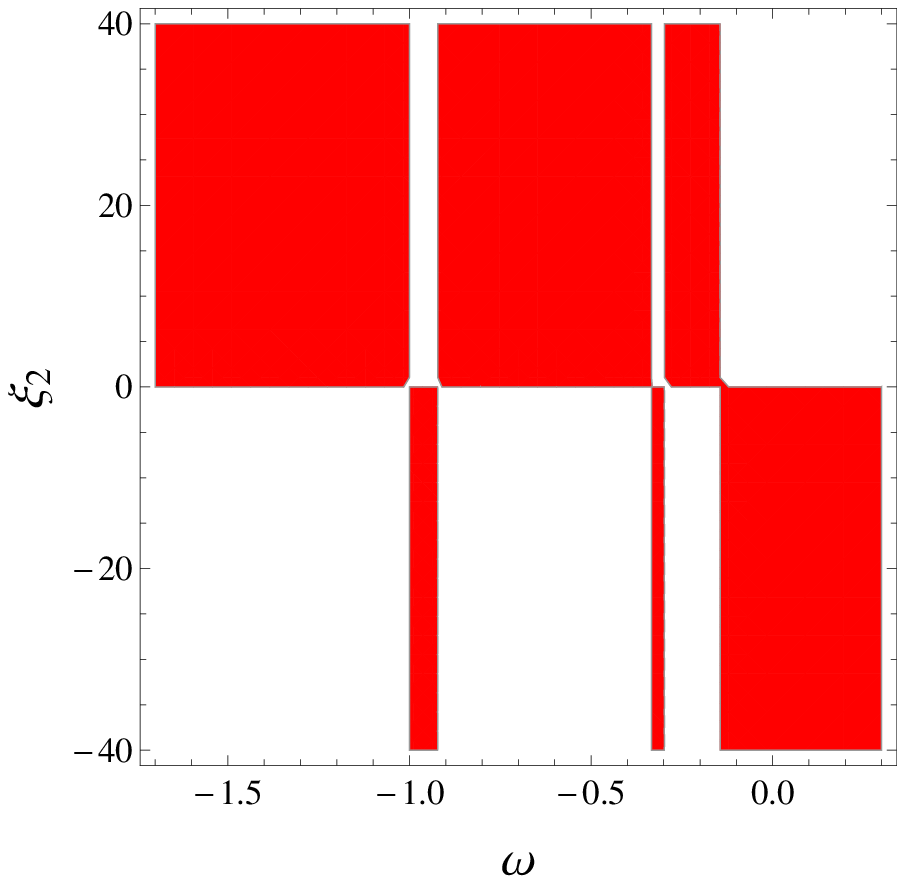,width=.5\linewidth}
\epsfig{file=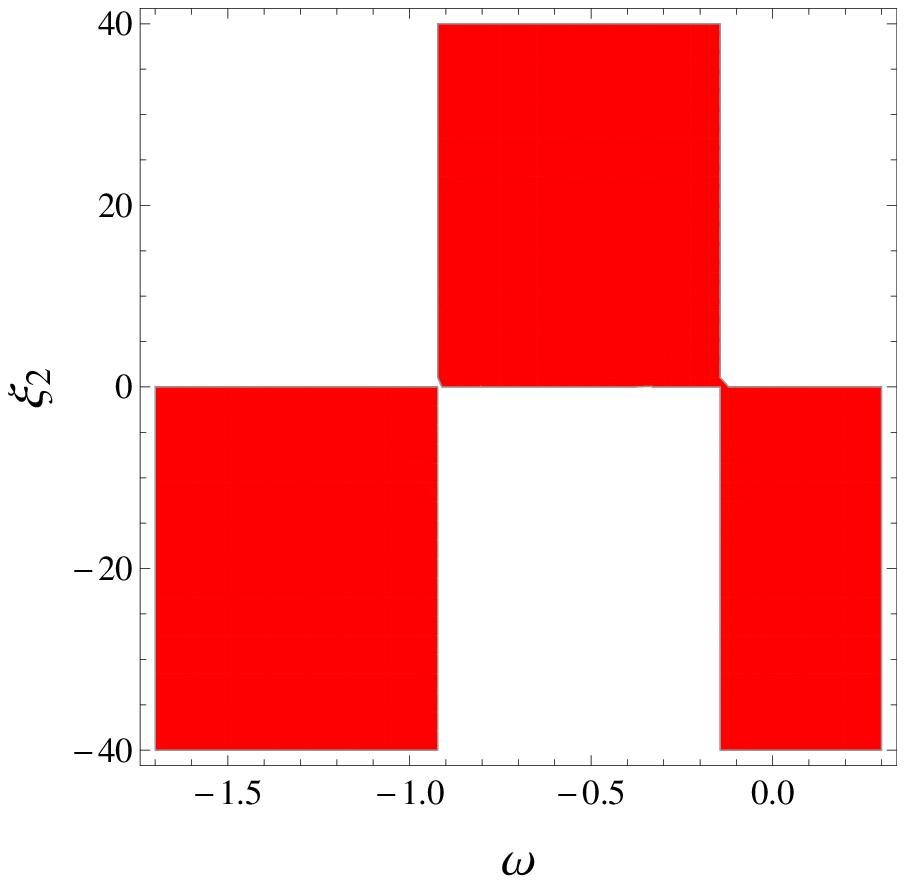,width=.5\linewidth} \caption{Stable modes in
$(\omega, \xi_{2})$ space corresponding to $n$=0.5,5 from left to
right.}
\end{figure}
\begin{figure}
\epsfig{file=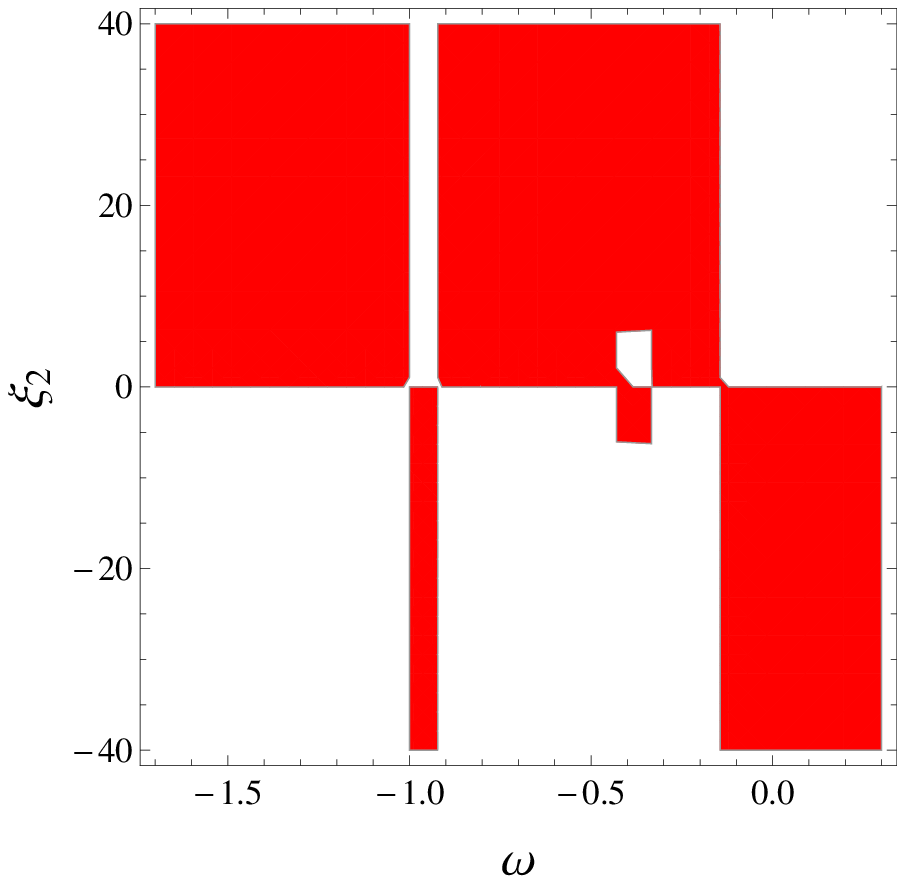,width=.5\linewidth}
\epsfig{file=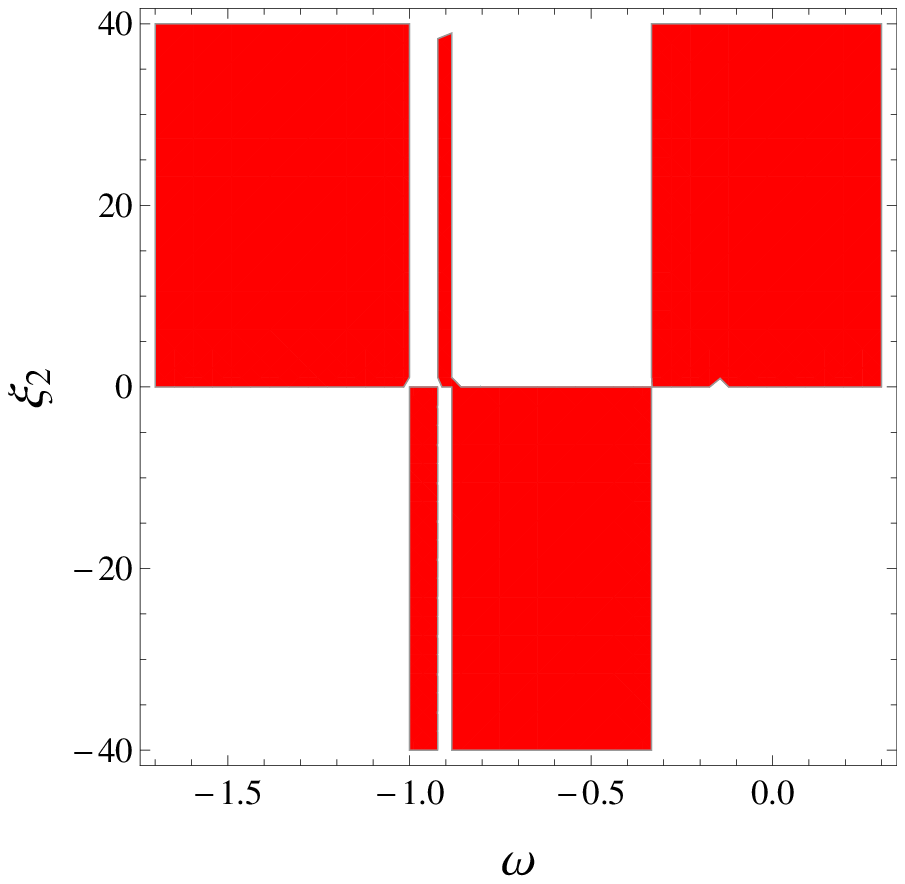,width=.5\linewidth}\caption{Stable regions in
$(\omega, \xi_{2})$ space corresponding to $n$=-.05,-5 from left to
right.}
\end{figure}

\section{Final Remarks}

The stability of the EU is considered the most debatable problem in
cosmology. Einstein tried to find a static solution of his field
equations to describe the isotropic and homogeneous universe. Since
the field equations of GR have no static solution, therefore,
Einstein introduced the term known as cosmological constant
$(\Lambda)$ to have static solutions. It is important to know
whether it can provide a natural initial state for a past eternal
universe, whether it allows the universe to evolve away from this
state, and whether under any circumstances it can act as an
attractor for the very early evolution of the universe. With these
questions in mind, we have investigated in detail that the closed EU
is stable or unstable against linear homogeneous perturbations.

One of the most fundamental issues in cosmology is the mystery
behind the beginning as well as the origin of the universe.
According to some physical constraints on cosmic matter
configuration, GR equations suggest that the current expanding
cosmos must be anticipated by a singularity known as the big-bang
singularity, where the physical parameters like energy density and
spacetime curvature diverge. The emergent universe scenario is based
on the pillars of the stable EU and is considered a favorable
approach in cosmology to resolve the captivating issue of primordial
singularity. The initial state of the universe in this framework is
the EU instead of primordial singularity and then smoothly evolves
to the rapid exponential inflationary era \cite{26}. This conjecture
implies that the initial cosmic epoch is the EU which enters into an
expanding posture that leads to the inflationary phase. The
phenomenon of EU is mainly manifested by the closed FLRW universe
with isotropic fluid and the cosmological constant. The most
essential characteristic of a successful emergent universe depends
upon the stable solutions of the EU for any type of perturbation.

In this paper, we have investigated the stability of closed EU by
using homogenous perturbations in EMSG. This newly proposed theory
is non-conserved due to the interaction between curvature and matter
parts. We have established equations of motion for the static and
perturbed system through a linear EoS parameter. We have chosen
specific $f(\mathcal{R},\mathbf{T}^{2})$ models to construct the
perturbed differential equations whose solutions help to explore the
stable regimes of closed EU. For the considered models, we have
examined the conserved/non-conserved EMT cases against the
homogeneous perturbations technique. The major results are given as
follows.
\begin{itemize}
\item We have formulated a particular expression of
$f_{2}(\mathbf{T}^{2})$ for conserved EMT which fulfills the
conservation equation. It is analyzed that stable regions of the
closed EU exist for all choices of the integration constant.
\item In the non-conserved case, we have considered a specific form of
$f_{2}(\mathbf{T}^{2})$ and examined the stability of closed EU for
arbitrary values of the model parameters. It is found that the
stable closed EU exists for all values of the model parameters.
\end{itemize}
We conclude that for all values of the EoS parameter, stable regions
of closed EU appear in EMSG. Since all initial vector perturbations
remain frozen, therefore, the stable modes by vector perturbations
also appear for all EoS values.

It is worthwhile to mention here that the range of $\omega$ in EMSG
increases and provides stable modes corresponding to all values of
$\omega$ for both conserved as well as non-conserved EMT that are
not stable in other modified theories. For example in
$f(G,\mathcal{T})$ theory, it is found that for positive values of
integration constant, stable EU exists for negative values of
$\omega$ while no stable region exists for its positive values as
well as negative values of integration constant in the conserved EMT
case. For non-conserved EMT, stable regions are observed only for
negative values of the model parameter and EoS parameter \cite{27}.
In $f(\mathcal{R},\mathcal{T},Q)$ theory, it is analyzed that for
positive values of integration constant, the stable EU exists only
for $\omega<1$ and stable modes exist in the range
$-1<\omega<\frac{1}{3}$ for negative values of integration constant
in the conserved EMT case. In the non-conserved EMT case, stability
decreases with decreasing value of the model parameter \cite{28}. It
would be interesting to extend this work for inhomogeneous
perturbations which indeed could yield a useful framework for the
stability of closed EU in EMSG.

\end{document}